\documentclass[preprint,showpacs,showkewords,preprintnumber,amsmath,amssymb]{revtex4}
 \usepackage{tipa}
 \usepackage{amsmath}
 \usepackage{txfonts}
 \usepackage{amssymb}
 \usepackage{graphicx,epsfig}
 \usepackage{bm}
 \begin{document}

 \title{A novel mathematical construct for the family of leptonic mixing patterns}
 \author{Shu-jun Rong}\email{rongshj@snut.edu.cn}

 \affiliation{Department of Physics, Shaanxi University of Technology, Hanzhong, Shaanxi 723000, China}

 \begin{abstract}
In order to induce a family of mixing patterns of leptons which
accommodate the experimental data with a simple mathematical construct, we construct a novel object from the hybrid of two elements of a finite group with a parameter $\theta$. This construct is an element of a mathematical structure called group-algebra. It could reduce to a generator of a cyclic group if $\theta/2\pi$ is a rational number. We discuss a specific example on the base of  the group $S_{4}$. This example shows that infinite cyclic groups could give the viable mixing patterns for Dirac neutrinos.

\end{abstract}


 \maketitle

 \section{Introduction}
How to explain the mixing pattern of leptons is a challenging question in neutrino physics. One of the most popular strategies is resorting to a discrete flavor group~\cite{1,2,3,4,5,6,7,8,9}. A special group $G_{f}$ is
chosen to design the Lagrangian of the general theory. Then after the spontaneous symmetry breaking of $G_{f}$, the residual symmetries in the leptonic sector could determine the mixing pattern of leptons~\cite{10}.
However, the mixing pattern accommodating the experimental data is not unique. There are a set of patterns which could not be discriminated by experiments. Furthermore, considering the results of the scan of finite groups, the order of the flavor group which gives a viable pattern is large~\cite{11,12,13}. Accordingly, the dynamic model based on these groups is complex in techniques. So we confront an important question whether there is a simple mathematical construct to induce a family of mixing patterns which satisfy the experimental constraints. In this paper, we introduce a mathematical object inspired by the idea in quantum mechanics (QM). As we know, a general state of a system in QM could be decomposed with the eigenstates of some observables . With the free coefficients of superposition, we could obtain a complete set of the states.
Similarly, we speculate that the mathematical construct which gives a family of mixing patterns could be composed of simple group elements which predict special patterns. Its expression is as follow
\begin{equation}
\label{eq:1}
X(\theta)=\cos\theta A + i\sin\theta B,
\end{equation}
where A, B are elements of a small finite group $G$ in the 3-dimensional unitary representation and $\theta$ is a parameter to mark the hybrid of them, $i$ is the imaginary factor. In modern mathematics, $X(\theta)$ is an element of the group-algebra $F[G]$ constructed by introduction of addition in the group $G$. To keep that $X(\theta)$ is unitary, A, B satisfy the condition
\begin{equation}
\label{eq:2}
AB^{+}=BA^{+},~~ A^{+}B=B^{+}A.
\end{equation}
This condition is equivalent to the following one
\begin{equation}
\label{eq:3}
A=C_{1}B,~~ A=BC_{2},~~ with~~ C^{2}_{1}=C^{2}_{2}=I,
\end{equation}
where $I$ is the identity matrix.

We suppose that the neutrino sector satisfies the
symmetry expressed by $X(\theta)$. So the mass matrix of neutrinos follows the relation
\begin{equation}
X^{T}(\theta)M_{\nu}X(\theta)=M_{\nu},
\end{equation}
 for Majorana neutrinos or
\begin{equation}
X^{+}(\theta)M^{+}_{\nu}M_{\nu}X(\theta)=M^{+}_{\nu}M_{\nu},
\end{equation}
for Dirac neutrinos.
Then employing the leptonic mixing matrix, $X(\theta)$ could be diagonalized as
\begin{equation}
U^{+}X(\theta)U=diag(\pm1,~~ \pm1,~~ \pm1),~~ for~~ Majorana~~ neutrinos,
\end{equation}
\begin{equation}
U^{+}X(\theta)U=diag(e^{i\alpha},~~ e^{i\beta},~~ e^{i\gamma}),~~  for~~ Dirac~~ neutrinos,
\end{equation}
where $U$ is the Pontecorvo-Maki-Nakagawa-Sakata (PMNS) mixing matrix. Here we work in the basis where the mass matrix of charged leptons
is diagonal. In other words, the symmetry in this sector is trivial which could only affect the mixing matrix through the permutation of rows and nonphysical phases.
We note that $X(\theta)$ is an element of the group $Z_{2}$ in the case of Majorana neutrinos. It has been widely studied in published literature. We focus on the case of Dirac neutrinos in this paper.
In the following section, we give an specific realisation of our construct.

\section{An example}
Once the group elements $A$, $B$ are given, we could obtain a set of mixing patterns with the eigenvectors of the symmetry $X(\theta)$. As an illustrative example, we choose
$A$, $B$ from the group $S_{4}$ generated by 3 elements which observe the following relations~\cite{14,15}:
\begin{equation}
S^{2}=V^{2}=(SV)^{2}=(TV)^{2}=E,~~
T^{3}=(ST)^{3}=E, ~~(STV)^{4}=E,
\end{equation}
where $E$ is the identity element. The 3-dimensional representations of the generators are expressed as~\cite{14,15}
\begin{equation}
\ S=\frac{1}{3}\left(
                 \begin{array}{ccc}
                   -1 & 2 & 2 \\
                   2 & -1 & 2 \\
                   2 & 2 & -1 \\
                 \end{array}
               \right),\ T=\left(
                 \begin{array}{ccc}
                   1 & 0 & 0 \\
                   0 & \omega^{2} & 0 \\
                   0 & 0 & \omega\\
                 \end{array}
               \right),\  V=\mp\left(
                 \begin{array}{ccc}
                   1 &0 & 0 \\
                   0 & 0 & 1 \\
                   0 & 1 & 0 \\
                 \end{array}
               \right),
\end{equation}
where $\omega=e^{i2\pi/3}$, the sign $\mp$ denotes the representation $\mathbf{3}$ and $\mathbf{3}^{\prime}$ respectively.
Considering the unitary condition in the form of Eq.~(\ref{eq:2}) or Eq.~(\ref{eq:3}), a viable realization of $A$, $B$ reads $A=TV$, $B=STV$. With the minus sign for the representation of $V$, $X(\theta)$
is expressed as
\begin{equation}
\label{eq:10}
X(\theta)=\left(
\begin{array}{ccc}
 \frac{i}{3} \sin \theta -\cos \theta & \frac{2}{3} e^{i \pi /6} \sin \theta  & -\frac{2}{3}e^{-i\pi/6} \sin \theta  \\
 -\frac{2i}{3} \sin \theta  & \frac{2}{3}e^{i \pi /6} \sin \theta & \frac{1}{3} e^{-i\pi/6}\sin \theta -e^{-i2\pi/3}  \cos \theta  \\
 -\frac{2i}{3} \sin \theta  & -\frac{1}{3} e^{i\pi/6}\sin \theta-e^{i 2\pi/3}\cos \theta  & -\frac{2}{3}e^{-i \pi /6} \sin \theta \\
\end{array}
\right).
\end{equation}
Note that the sign of $V$ is not important in this case. The replacement $V\rightarrow-V$ is equivalent to the transformation $X(\theta)\rightarrow-X(\theta)$.
The eigenvalues and corresponding eigenvectors of $X(\theta)$ are
\begin{equation}
\lambda_{1}=-1, ~~\lambda_{2}=-e^{-i\theta},~~ \lambda_{3}=1,
\end{equation}
 \begin{equation}
u_{1}=\frac{1}{N_{1}}\left(
         \begin{array}{c}
           \frac{e^{i\theta}+1}{e^{i\theta}+1-\sqrt{3}e^{-i\pi/6}} \\
           \frac{e^{i\pi/3}-e^{i\theta}}{e^{-i\pi/3}-e^{i\theta}} \\
           1 \\
         \end{array}
       \right),~~u_{2}=\frac{1}{\sqrt{3}}\left(
                         \begin{array}{c}
                           \omega\\
                           \omega^{2} \\
                           1\\
                         \end{array}
                       \right),~~u_{3}=\frac{1}{N_{3}}\left(
         \begin{array}{c}
           \frac{e^{i\theta}-1}{e^{i\theta}-1+\sqrt{3}e^{-i\pi/6}} \\
           \frac{e^{i\pi/3}+e^{i\theta}}{e^{-i\pi/3}+e^{i\theta}} \\
           1 \\
         \end{array}
       \right),
\end{equation}
 where $N_{1}$ and $N_{3}$ are normalization factors. Then the mixing matrix of leptons reads
 \begin{equation}
 \label{eq:13}
 U(\theta) =(u_{1}~~ u_{2}~~ u_{3})
 \end{equation}
up to permutations of rows or columns and nonphysical phases.
We give some comments here:\\
1. Although $X(\theta)$ could be transformed to the diagonal form, i.e., $diag(-1, ~e^{-i\theta},~1)$,  in general, it is not equivalent to a $U(1)$ group because $U(\theta)$
depends on the parameter $\theta$. Furthermore, $X(\theta)$ could not denotes an infinite group in general cases, because neither the identity element nor the inverse of $X(\theta)$ could be written in the form of
$X(\theta^{\prime})$ .\\
2. If $\theta/2\pi $ equals $\pm j/k$, where natural number $i, j$ are coprime , $X(\theta)$ reduces to a generator of the cyclic group $Z_{2k}$ or $Z_{k}$ when $k$ is odd or even.
Furthermore, we note that if a special value, i.e., $2(\pm j/k)\pi$ of $\theta$, could make the mixing matrix $U(\theta)$
accommodate the experimental data, infinite ones in the form  $2(\pm j/k\pm j^{\prime}/k^{\prime})\pi$ could also hold when $j^{\prime}/k^{\prime}$ is small enough. In other words, if a finite group survives in the experimental constraints, so do infinite ones.\\
3. Because the translation $\theta\rightarrow\theta+\pi$ is equivalent to the replacement $X(\theta)\rightarrow-X(\theta)$, so the independent range of the parameter $\theta$ is $[0, \pi]$.

Now we give the phenomenological results of $X(\theta)$.
Compared with the standard parametrization of the mixing matrix, i.e.,
\begin{equation}
U=
\left(
\begin{array}{ccc}
 c_{12}c_{13} & s_{12}c_{13} & s_{13}e^{-i\delta} \\
 -s_{12}c_{23}-c_{12}s_{13}s_{23}e^{i\delta} & c_{12}c_{23}-s_{12}s_{13}s_{23}e^{i\delta} & c_{13}s_{23} \\
s_{12}s_{23}-c_{12}s_{13}c_{23}e^{i\delta} & -c_{12}s_{23}-s_{12}s_{13}c_{23}e^{i\delta} & c_{13}c_{23}
\end{array}
\right),
\end{equation}
where $s_{ij}\equiv\sin{\theta_{ij}}$, $c_{ij}\equiv\cos{\theta_{ij}}$, $\delta$ is the Dirac CP-violating phase,
we could obtain the mixing angles with the following relations:
\begin{equation}
\sin^{2}\theta_{13}=|U_{e3}|^{2}, ~~\sin^{2}\theta_{23}=|U_{\mu3}|^{2}/(1-|U_{e3}|^{2}),~~\sin^{2}\theta_{12}=|U_{e2}|^{2}/(1-|U_{e3}|^{2}).
\end{equation}
Employing the $\chi^{2}$ function defined as
\begin{equation}
\chi^{2}=\sum_{ij=13,23,12}(\frac{\sin^{2}\theta_{ij}-(\sin^{2}\theta_{ij})_{exp}}{\sigma_{ij}})^{2},
\end{equation}
where $(\sin^{2}\theta_{ij})_{exp}$ are best global fit values from Ref.~\cite{16}, $\sigma_{ij}$ are 1$\sigma$ errors,
the best fit data of the parameter $\theta$ and mixing angles are shown in Table~\ref{tab:1}.
\begin{table}
\caption{Best fit data of the parameter $\theta$ and $\sin{\theta_{ij}}$ }
\label{tab:1}       % Give a unique label
% For LaTeX tables use
\begin{tabular}{|c|c|c|c|c|c|}
\noalign{\smallskip}\hline
 ~~$Ordering$~~&~~~~~$\chi^{2}_{min}~~~~~$ &~~~~~$\theta_{bf1}$~~~~~&~~$(\sin^{2}\theta_{13})_{bf}$~~& ~~$(\sin^{2}\theta_{23})_{bf}$~~&~~$(\sin^{2}\theta_{12})_{bf}$~~ \\[0.5ex]\hline
\noalign{\smallskip}\noalign{\smallskip}\hline
 ~~ Normal~~ &~11.9396~&~~0.11496$\pi$~~& ~0.021504~&~~0.39575~~&~~0.34066~~\\
\hline
~~Inverted~~&~~9.0379~~&~~0.11546$\pi$~~~&~~0.02169~~& ~0.6047~&~~0.34072~~\\
\hline
\noalign{\smallskip}
\end{tabular}
% Or use
\vspace*{0.5cm}  % with the correct table height
\end{table}
Note that because of the freedom of permutations of rows or columns of the mixing matrix in Eq.~(\ref{eq:13}), the best fit value of $\theta$ in the range [0,~$\pi$] is not unique.
The complete set of the best fit values for either mass ordering is listed as
\begin{equation}
\begin{array}{c}
  \theta_{bf1},~\theta_{bf2}=\pi/3-\theta_{bf1}, ~\theta_{bf3}=\pi/3+\theta_{bf1},  \\
 \theta_{bf4}=2\pi/3-\theta_{bf1}, ~\theta_{bf5}=2\pi/3+\theta_{bf1}, \theta_{bf6}=\pi-\theta_{bf1}.
\end{array}
\end{equation}
In the case of normal mass ordering, these best fit data correspond to the following mixing matrix respectively:
\begin{equation}
U_{1}=S_{23}U,~U_{2}=S_{23}S_{12}US_{13},~U_{3}=S_{12}US_{13},~U_{4}=S_{13}S_{12}U,~U_{5}=S_{13}U,~U_{6}=US_{13},
\end{equation}
where $U$ is written as Eq~(\ref{eq:13}), the permutation matrices $S_{ij}$ are expressed as
\begin{equation}
S_{23}=\left(
         \begin{array}{ccc}
           1 & 0 & 0 \\
           0 & 0 & 1 \\
           0 & 1 & 0 \\
         \end{array}
       \right), ~~S_{13}=\left(
         \begin{array}{ccc}
           0 & 0 & 1 \\
           0 & 1 & 0 \\
           1 & 0 & 0 \\
         \end{array}
       \right),~~S_{12}=\left(
         \begin{array}{ccc}
           0 & 1 & 0 \\
           1 & 0 & 0 \\
           0 & 0 & 1 \\
         \end{array}
       \right).
\end{equation}
The mixing matrices in the case of inverted mass ordering are $U^{\prime}_{j}=S_{23}U_{j}$, with $j=1,~2,~3,~4,~5,~6$.
Either in the case of normal or inverted mass ordering, the six  best fit values of $\theta$ correspond to the same $\chi^{2}_{min}$ and the same mixing angles. This observation could seen from the function $\chi^{2}$ shown in Figure~\ref{fig:1}.
And some comments are given as follows:\\
1. Begin with the curve of the function $\chi^{2}$ of $U_{1}$ or $U^{\prime}_{1}$, other curves of $\chi^{2}$ could be obtained by the space inversion with the axis $\theta=k\pi/6$ step by step, with $k=1,~2,~3,~4,~5.$
Accordingly, the range of the parameter $\theta$ at the level of $3\sigma$ of any mixing matrix could be derived from that of a special mixing matrix. For example, the range of $\theta$ at $3\sigma$ level for $U_{3}$
and $U^{\prime}_{3}$ is [$0.4422\pi, ~0.4545\pi$],~[$0.4422\pi, ~0.4546\pi$] respectively. So that for  $U_{2}$
and $U^{\prime}_{2}$ is [$(2/3-0.4545)\pi, ~(2/3-0.4522)\pi$],~[$(2/3-0.4546)\pi, ~(2/3-0.4522)\pi$] respectively. The range of $\theta$ at $3\sigma$ level for any mixing matrix is narrow, which results from the stringent experimental constraint of $\sin\theta_{13}$. This observation could seen from the the correlation of two mixing angels such as those of $U_{3}$. See Figure ~\ref{fig:2} for example, where the mixing angle $\theta_{13}$
covers its full range at $3\sigma$ level, while $\theta_{23}$ or $\theta_{12}$  covers only several percents of the measure of its $3\sigma$ range.\\
2.  As the parameter $\theta$ varies in the range at $3\sigma$ level, we could obtain a family of mixing patterns. Specially, the values which could be expressed as $2(j/k)\pi$ are in the range of $3\sigma$ level  for any mixing matrix. For example, $4\pi/9$ is in the range of $\theta$ for the matrix $U_{3}$ or $U^{\prime}_{3}$. $X(\theta=4\pi/9)$ is a generator of the group $Z_{18}$. We note that $Z_{18}$ have been obtained as a residual symmetry in the neutrino sector by the scan of thousands of finite groups in the literature \cite{12}. According to the aforementioned observation, there are infinite cyclic groups $Z_{n}$ which could accommodate the experimental data. Furthermore, as an interesting phenomenon, the value $\theta=\pi/3$ which is a axis of the reflection of two curves of $\chi^{2}$ corresponds to the well known tri-bi-maximal (TBM) mixing pattern~\cite{17,18}, i.e., $U(\theta=\pi/3)=U_{TBM}$. \\
3. The Dirac CP phase could obtained through the Jarlskog invariant expressed as~\cite{19}
\begin{equation}
J_{cp}\equiv\mathrm{Im}[U_{22}U^{*}_{23}U^{*}_{32}U_{33}]=\frac{1}{8}\sin2\theta_{13}\sin2\theta_{23}\sin2\theta_{12}\cos\theta_{13}\sin\delta.
\end{equation}
Substitute $J_{CP}$ with the elements of $U(\theta)$, we find it equals 0 independent of the parameter $\theta$. Hence the Dirac CP phase is trivial, i.e., $\sin\delta=0$.

\begin{figure}[tbp]
\label{fig:1}
\centering % \begin{center}/\end{center} takes some additional vertical space
\includegraphics[width=.48\textwidth]{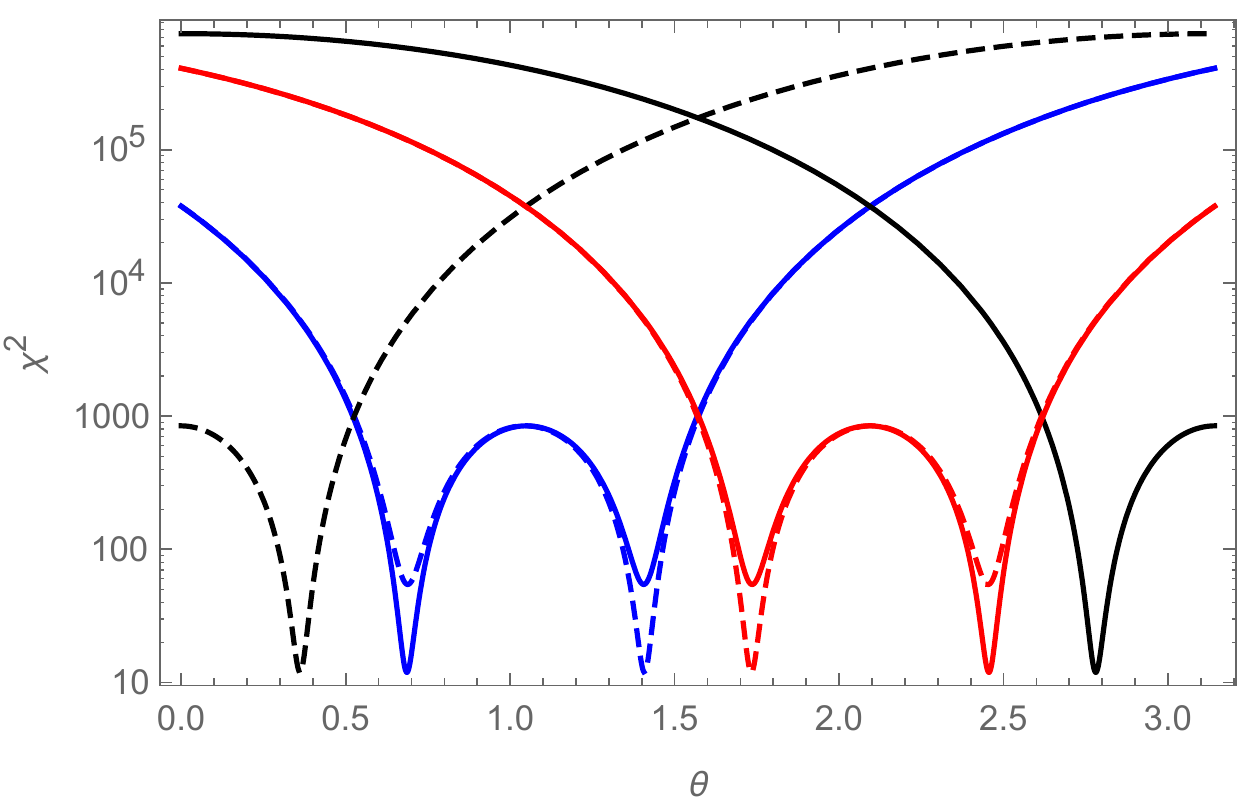}
\hfill
\includegraphics[width=.48\textwidth]{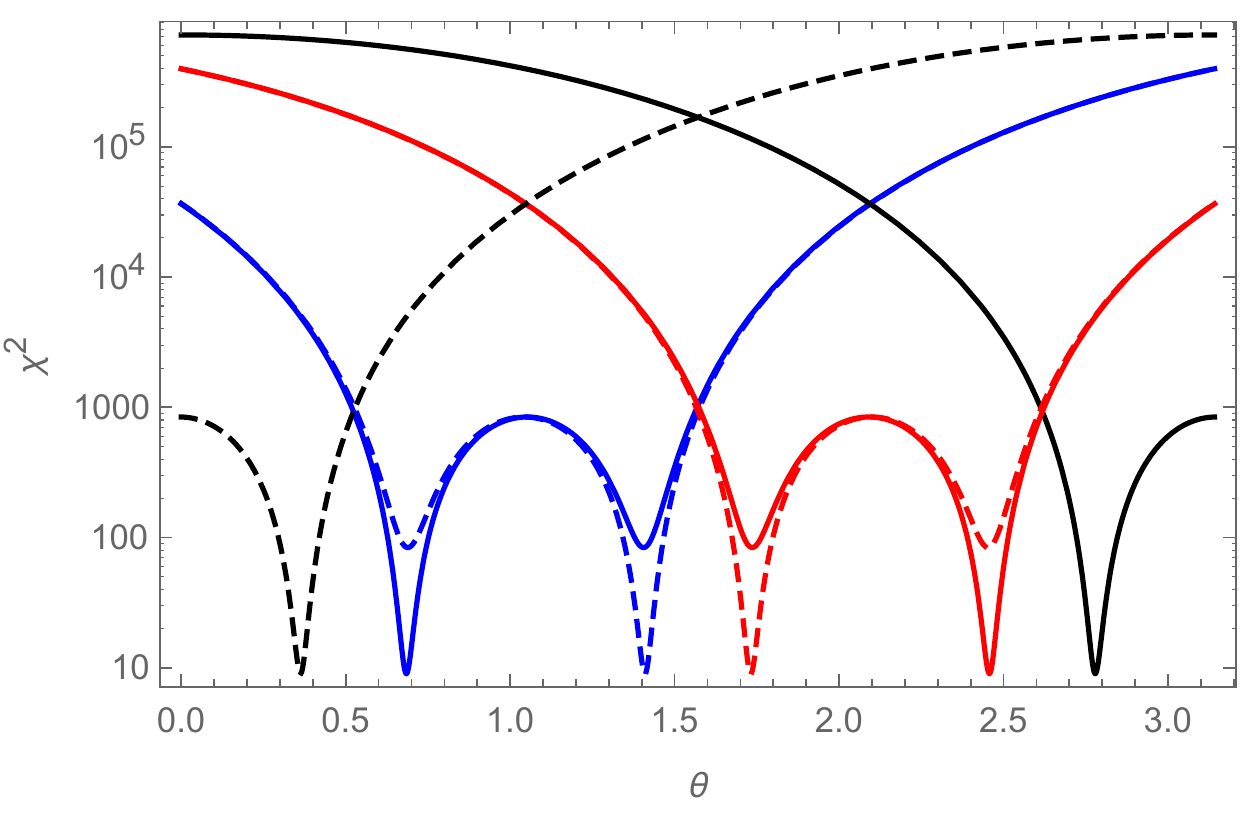}
\hfill
\caption{\label{fig:1} The function $\chi^{2}$ for the viable  mixing matrices. The left panel is for the case of normal mass ordering and the right one is for the inverted case. The black dashed curve is the function $\chi^{2}$ of $U_{1}(U^{\prime}_{1})$, the blue curve is for $\chi^{2}$ of $U_{2}(U^{\prime}_{2})$, the blue dashed curve is for $\chi^{2}$ of $U_{3}(U^{\prime}_{3})$, the red dashed curve is for $\chi^{2}$ of $U_{4}(U^{\prime}_{4})$, the red curve is for $\chi^{2}$ of $U_{5}(U^{\prime}_{5})$, the black curve is for $\chi^{2}$ of $U_{6}(U^{\prime}_{6})$.}
\end{figure}

\begin{figure}[tbp]
\label{fig:2}
\centering % \begin{center}/\end{center} takes some additional vertical space
\includegraphics[width=.48\textwidth]{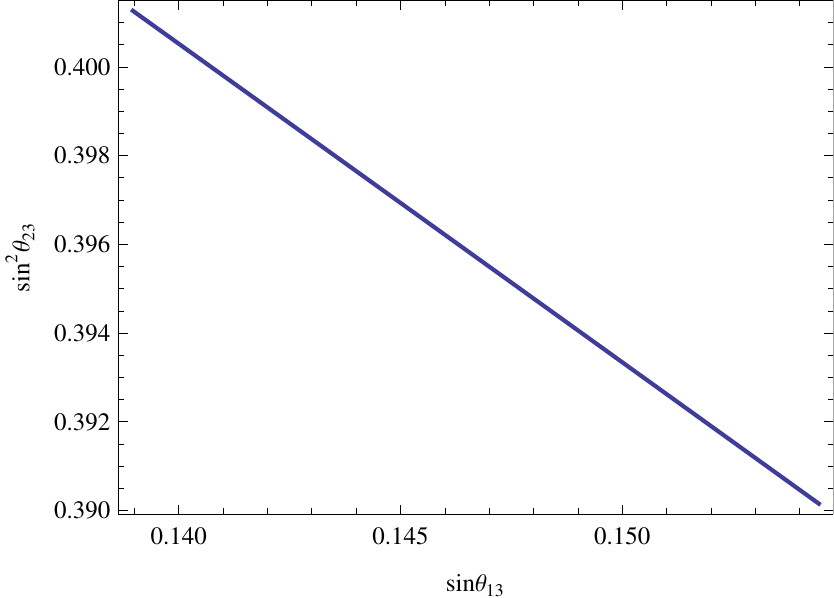}
\hfill
\includegraphics[width=.48\textwidth]{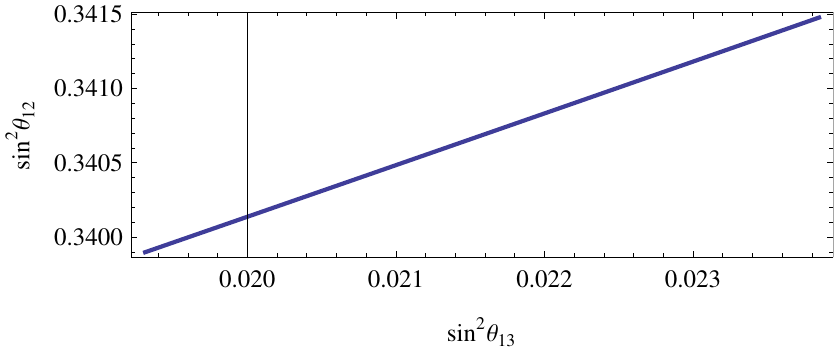}
\hfill
\caption{\label{fig:2} Correlations of the mixing angles of the mixing matrix $U_{3}$.}
\end{figure}

\section{Conclusion}
Inspired by the ideal in QM, as a generalization of application of finite groups, we construct a mathematical structure called group-algebra to give a family of mixing patterns of leptons.
This construct with a parameter $\theta$ could reduce to a generator of a cyclic group when $\theta/2\pi$ is a rational number.
A specific example made from the group $S_{4}$ is given, which shows that a set of mixing patterns for Dirac neutrinos could accommodate the experimental data at the $3\sigma$ level.
And infinite cyclic groups could work as the symmetry in the neutrino sector, which supplements the observation where several ones are found by the scan of finite groups up to order 2000 in the literature.
Construct of a dynamical model on the base of this mathematical structure is intriguing, which will be considered in our future work.

\acknowledgments
 This work was supported by the National Natural Science Foundation of China under the Grant No. 11405101 and the research foundation of Shaanxi University of Technology under the Grant No. SLGQD-13-10.
The author declares that there is no conflict of interest regarding the publication of this paper.

\end{document}